\documentclass[twocolumn]{aastex631}
\usepackage{amsmath}
\usepackage{wrapfig}
\usepackage{svg}
\usepackage{array}
\usepackage{booktabs}
\usepackage{multirow} 

\newcommand{\Mbh}{M_{\rm BH}}
\newcommand{\mbh}{m_\bullet}
\newcommand{\mstar}{m_\star}
\newcommand{\Nbh}{N_\bullet}
\newcommand{\nbh}{n_\bullet}
\newcommand{\Nstar}{N_\star}
\newcommand{\nstar}{n_\star}
\newcommand{\msb}{\frac{\mstar}{\mbh}}
\newcommand{\Msun}{M_{\odot}}
\newcommand{\TL}{T_{\rm LISA}}

\newcommand{\fbh}{\mathit{f}_{\bullet}}
\newcommand{\msbp}[1]{\left(\msb\right)^{#1}}

\newcommand{\RI}{R_{\rm I}} 
\newcommand{\RII}{R_{\rm II}}
\newcommand{\lnL}{\log{\Lambda}}
\newcommand{\RGW}{R_{\rm GW}}

\newcommand{\fobs}{f_{\rm obs}}
\newcommand{\Mcobs}{\mathcal{M}_{\rm obs}}
\newcommand{\Mobs}{M_{\rm obs}}
\newcommand{\Mmw}{M_{\rm MW}}
\newcommand{\dd}{{\rm d}}

\newcolumntype{L}[1]{>{\raggedright\arraybackslash}p{#1}} 

\def\approxprop{%
  \def\p{%
    \setbox0=\vbox{\hbox{$\propto$}}%
    \ht0=0.6ex \box0 }%
  \def\s{%
    \vbox{\hbox{$\sim$}}%
  }%
  \mathrel{\raisebox{0.7ex}{%
      \mbox{$\underset{\s}{\p}$}%
    }}%
}

\begin{document}

\title{Dynamics around supermassive black holes: \\ Extreme-mass-ratio inspirals as Gravitational-wave Sources}

\correspondingauthor{Barak Rom}
\email{barak.rom@mail.huji.ac.il}

\author[0000-0002-7420-3578]{Barak Rom}
\affiliation{Racah Institute of Physics, The Hebrew University of Jerusalem, 9190401, Israel}

\author[0000-0002-8304-1988]{Itai Linial}
\affil{Department of Physics and Columbia Astrophysics Laboratory, Columbia University, New York, NY 10027, USA}
\affil{Institute for Advanced Study, 1 Einstein Drive, Princeton, NJ 08540, USA}

\author[0000-0001-6474-4402]{Karamveer Kaur}
\affiliation{Technion - Israel Institute of Technology, Haifa, 3200002, Israel}
\affiliation{Racah Institute of Physics, The Hebrew University of Jerusalem, 9190401, Israel}

\author[0000-0002-1084-3656]{Re'em Sari}
\affiliation{Racah Institute of Physics, The Hebrew University of Jerusalem, 9190401, Israel}

\shorttitle{EMRIs as LISA Sources}
\shortauthors{Rom, Linial, Kaur \& Sari}

\begin{abstract}
\noindent
Supermassive black holes and their surrounding dense stellar environments nourish a variety of astrophysical phenomena. 
We focus on the distribution of stellar-mass black holes around the supermassive black hole and the consequent formation of extreme-mass-ratio inspirals (EMRIs). 
We derive a steady-state distribution, considering the effects of two-body scattering and gravitational-wave emission, and calculate the EMRI formation rate, eccentricity distribution, and EMRI-to-plunge ratio.
Our model predicts: {\bf (a)} a stronger segregation than previously estimated at the outskirts of the sphere of influence (at $\sim0.01-2\rm pc$ for a Milky Way-like galaxy); 
{\bf (b)} an increased EMRI-to-plunge ratio, favoring EMRIs at galaxies where stellar-mass black holes are scarce;
{\bf (c)} a detection of about $2\times10^3$ resolvable EMRIs, with a signal-to-noise ratio above $20$, along a $4\ \rm yr$ LISA mission time;
and {\bf (d)} a confusion noise, induced by a cosmological population of unresolved EMRIs, reducing the LISA sensitivity in the $1-5\ \rm mHz$ frequency range by up to a factor of $\approx2$, relative to the instrumental noise.
\end{abstract}

\keywords{Galactic center (565), Stellar dynamics (1596), Supermassive black holes (1663), Stellar mass black holes(1611), Gravitational wave sources(677), Gravitational wave detectors(676)}


\section{Introduction}
Supermassive black holes (SMBHs) and their surrounding nuclear stellar clusters (NSCs) reside at the centers of galaxies and cultivate a plethora of astrophysical phenomena: from quasars and active galactic nuclei \citep{Schmidt_1963,Netzer_2015} to tidal disruption events \citep[TDEs;][]{Rees_88,Gezari_21} and quasiperiodic eruptions \citep[QPEs;][]{Miniutti_2019,Arcodia_2021}.
Furthermore, the dense stellar environment in galactic nuclei is a fertile ground for gravitational-wave (GW) sources, including merging stellar-mass black hole (sBH) binaries \citep{Mapelli_2021,sedda_2023}, observable by the LIGO-Virgo-KAGRA collaboration, and mergers of sBHs with the SMBH, i.e., extreme-mass-ratio inspirals (EMRIs). The latter are expected to be primary GW sources for the space-based observatory LISA \citep{Amaro_2023}, with an expected detection rate of a few to thousands of EMRIs per year \citep{Gair_2004,Mapelli_2012,Babak_17,Bonetti_2020,Pozzoli_2023}. 
The large span in the estimated detection rate is mostly due to the uncertainties regarding the SMBH mass function and the EMRI rate per galaxy, ranging between $\sim10$ and $10^3 \rm\ Gyr^{-1}$ \citep{Hils_1995,Sigurdsson_1997,Ivanov_2002,Hopman_2005,Amaro_Seoane_2011,Merritt_2015,Aharon_2016,Bar_or_2016,Aceves_2022,Broggi_2022}.

The EMRI rate per galaxy depends on the stellar dynamics in the NSC, which has been extensively studied for over half a century.
\cite{BW_76} derived a zero-flux steady-state solution for a single-mass population, under the assumptions of spatial spherical symmetry with isotropic velocities, where the dynamics are dominated by weak two-body scattering. 
In this case, the phase-space distribution is given by $f(E)\propto E^p$ with $p=1/4$, and so the number density scales as $n(r)\propto r^{-\gamma}$, with $\gamma=3/2+p=7/4$ (hereafter, the BW profile). This solution satisfies a vanishing particle flux and a constant energy flux \citep{Rom_2023}. In a following paper, \cite{BW_77} generalized their calculation for multimass groups. Assuming that the most massive objects, with mass $m_{\max}$, are the most abundant, a zero-flux solution is satisfied when the massive group follows the single-mass BW profile, with $p_{\max}=1/4$, while the lighter objects, with mass $m_i<m_{\max}$, obtain shallower profiles, $p_i/p_{\max}\approx m_i/m_{\max}$ \citep{BW_77,Linial_2022}. 

However, when considering realistic NSCs, the light stars typically occur in much greater numbers than heavier compact objects. This led \cite{AH_09} to derive the ``strong mass segregation" solution \citep[see, for example,][]{Amaro_Seoane_2011,Aharon_2016,Amaro_2018}, which takes into account the drift of the massive objects toward the center of the cluster due to dynamical friction. The solution of \cite{AH_09} predicts that the light stars obtain the single-mass BW profile while the heavier objects follow a steeper profile, with $p=5/4$, corresponding to a nonzero constant inward flux. This solution was generalized for a continuous mass function by \cite{Keshet_2009}.

Recently, \cite{Linial_2022} have derived a steady-state solution for a continuous mass function that takes into account the dominance of different mass groups in different energy bins. Thus, this solution provides a self-consistent zero-flux solution in all energy bins and for each mass group simultaneously.

The steady-state distributions of the sBHs and stars affect the formation rates of the center-of-galaxy associated transients, such as TDEs, QPEs, and EMRIs, mainly through the interplay between two-body scattering and GW emission \citep{Alexander_2017,Amaro_2023}. 
Qualitatively, two-body scattering leads to a diffusion in angular momentum, producing highly eccentric orbits that efficiently dissipate energy by GW emission \citep[e.g.,][]{Hopman_2005,Hopman_2009,Amaro_Seoane_2011,Aharon_2016,Sari_Fragione_2019,Linial_2023}. 

In this work, we apply the general solution of \cite{Linial_2022} to a simplified model of a two-mass NSC, including solar-mass stars and sBHs. We introduce modifications to the steady-state distributions due to GW emission and derive analytically the characteristics of the resulting EMRIs. In Section \ref{sec:2}, we derive the spatial distribution of the sBHs. In Section \ref{sec:3}, we calculate the EMRI rate per galaxy, their eccentricity distribution, and the EMRI-to-plunge ratio. In Section \ref{sec:4} we estimate the expected number of EMRIs that will be detected by LISA and the residual GW background (GWB) from unresolved EMRIs. In Section \ref{sec:comp} we compare our model with known results in the literature. Finally, we summarize our results in Section \ref{sec:sum}.

\begin{deluxetable}{|c|c|l|} 
\tablewidth{1.0\columnwidth} 
\tablecaption{Distances Glossary \label{tab:glossary}}
\tablehead{
\colhead{Symbol} & \colhead{Eq.} & \colhead{Definition}
}
\startdata
$R_h$ & \ref{eq:Rh} & Radius of influence of the SMBH. \\ \hline
$R_s$ & \ref{eq:Rs} & Schwarzschild radius of the SMBH. \\ \hline
{ } & { } & Effective loss-cone due to GW emission - the\\ 
$r_{p, \rm lc}(r)$ & \ref{eq:2BGW} & periapsis below which GW dominates, for a \\ 
{ } & { } & given semimajor axis $r$.\\  \hline
$R_c$ & \ref{eq:Rc} & Distance differentiating plunge ($r\gtrsim R_c$) and \\ 
{  } & {  } & EMRI ($r\lesssim R_c$) progenitors. \\ \hline
$\RI$ & \ref{eq:RI} & Characteristic distance below which sBHs are\\ 
{  } & {  } & the dominant scatterers. \\ \hline
$\RII$ & \ref{eq:RII} & Characteristic distance replenished by sBHs \\ 
{  } & {  } & initially at $\RI$, defined as $\RII=r_{p,lc}(\RI)$. \\ \hline
{  } & {  } & Transition from scattering-dominated to \\ 
$\RGW$ & \ref{eq:RGW} & GW-dominated dynamics for circular orbits,  \\ 
{  } & {  } & defined as $\RGW=r_{p,\rm lc}(\RGW)$. \\ \hline
$R_t$ & \ref{eq:Rt} & Stellar tidal radius. \\ 
\enddata
\end{deluxetable}

\section{Steady-state Distributions in NSCs} \label{sec:2}

We study the distribution of stars and sBHs in the NSC that surrounds an SMBH of mass $\Mbh$. We assume that the cluster consists of solar-mass stars, $\mstar=M_{\odot}$, and sBHs, with $\mbh=10M_{\odot}$. Key scale radii are summarized in Table \ref{tab:glossary}. 
We focus on the dynamics within the radius of influence $R_h$, where the SMBH dominates the gravitational potential. Using the $M\propto\sigma_{h}^{\beta}$ relation, where $\sigma_{h}$ is the stellar velocity dispersion and $\beta=4$ \citep{Kormendy_2013}, the radius of influence is given by
\begin{equation} \label{eq:Rh}
R_h=\frac{G\Mbh}{\sigma_h^2}\simeq2\ {\rm pc}\left(\frac{\Mbh}{\Mmw}\right)^s,
\end{equation}
where $s=1-2/\beta=1/2$. We normalize $\Mbh$ to the mass of Sgr A*, $\Mmw= 4\times10^6 M_\odot$ \citep{Ghez_2008,Gillessen_2009}.

The steady-state distributions of stars and sBHs are determined by two-body scattering, which dominates at large distances from the SMBH, and dissipation due to GW emission, which takes over at the vicinity of the SMBH. 
 
Two-body scattering introduces a characteristic timescale for the angular momentum of a given orbit to change by order of itself: 
\begin{equation} \label{eq:t2b}
\tau^{(J)}_{2B}\left(r,r_p\right)=\frac{3\sqrt{2}\pi\left(3-\gamma\right)}{32c_{7/4}\lnL}\frac{P(r)}{\Nbh(r)}\left(\frac{\Mbh}{\mbh}\right)^2\frac{r_p}{r},
\end{equation}
where $r$ is the semimajor axis, $r_p$ is the periapsis, $P(r)$ is the orbital period, 
$\lnL$ is the Coulomb logarithm, with $\Lambda\approx\Mbh/\mstar$ (typically $\lnL\sim10$), and $\Nbh(r)$ is roughly the number of sBHs with semimajor axis between $r/2$ and $r$, defined as 
\begin{equation}\label{eq:Nn}
    \Nbh(r)=4\pi r^3 \nbh(r),
\end{equation}
where $\nbh(r)\propto r^{-\gamma}$ is the sBH number density.
The numerical prefactor in Eq. (\ref{eq:t2b}) is determined by the orbit-averaged diffusion coefficient \citep{Merritt_etal_2010}, with $c_{\gamma}$ depending on the scatterer number density, for a BW profile $c_{7/4}=1.35$ \citep{Bortolas_2019}.
Note that in Eq. (\ref{eq:t2b}) we assume that the sBHs dominate the scattering, which is valid at distances of order $r/R_h\lesssim 10^{-2}$, as discussed below. 

The GW characteristic energy-loss timescale, $E/\dot{E}$, for sBHs on highly eccentric orbits\footnote{A GR correction to the GW timescale, Eq. (\ref{eq:tgw}), was introduced by \cite{Zwick_2021}. However, for a nonspinning SMBH, it yields an order unity correction to the formation rate \citep[see][]{Aceves_2022} and therefore we neglect it here.} is given by \citep{Peters_1964}:
\begin{equation} \label{eq:tgw}
    \tau^{\rm (E)}_{\rm GW}\left(r,r_p\right)=\frac{96\sqrt{2}}{85}\frac{R_s}{c}\frac{\Mbh}{\mbh}\left(\frac{r_p}{R_s}\right)^4\sqrt{\frac{r}{r_p}},
\end{equation}
where $R_s$ is the Schwarzschild radius of the SMBH, given by
\begin{equation}\label{eq:Rs}
    R_s=\frac{2G\Mbh}{c^2}\simeq 4\times10^{-7}\ {\rm pc}\left(\frac{\Mbh}{\Mmw}\right).
\end{equation}

\begin{figure}[!ht] 
    \centering
    \includegraphics[width=8.6cm]{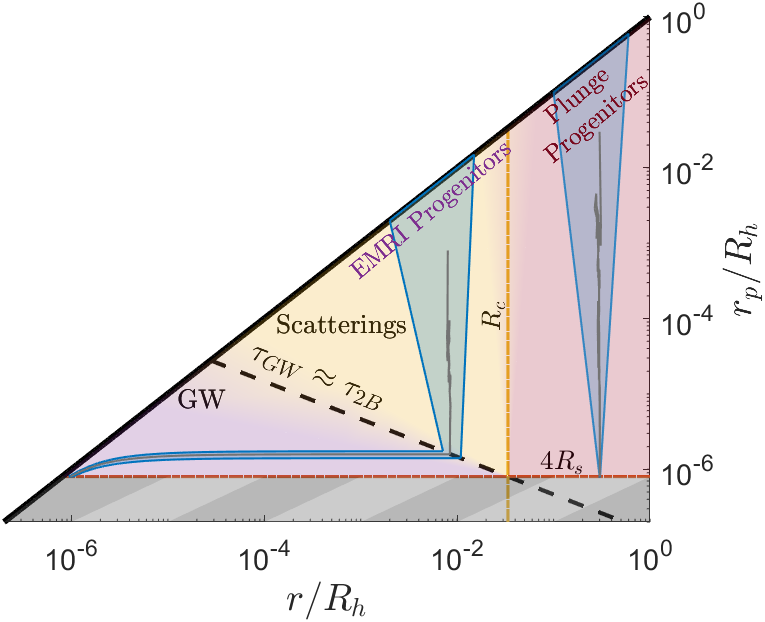}
    \caption{Orbital dynamics in the $(r,r_p)$ phase space, where $r$ is the semimajor axis and $r_p$ is the periapsis. The dashed black line distinguishes between the two-body-scattering-dominated region and the GW-dominated one (Eq. \ref{eq:2BGW}); above it, the scattering leads to a diffusion dominantly in the $r_p$ direction (as encapsulated by the shaded vertical blue funnel), while below it, GW emission efficiently shrinks and circularizes the orbit (along the horizontal shaded funnel). The red dashed-dotted line marks the mostly bound orbit, with $r_p=4R_s$, and the yellow dashed-dotted line represents the critical semimajor axis $R_c$ (Eq. \ref{eq:Rc}), separating EMRI progenitors (with $r\lesssim R_c$) and plunge progenitors (with $r\gtrsim R_c$). The gray lines demonstrate examples of trajectories leading to an EMRI or a plunge, according to their initial semimajor axis.}
    \label{fig:1}
\end{figure}

The transition between the scattering-dominated region and the GW-dominated one occurs when the two-body scattering timescale (Eq. \ref{eq:t2b}) and the GW timescale (Eq. \ref{eq:tgw}) are comparable \citep{Hopman_2005,Amaro_2018,Sari_Fragione_2019}
\begin{equation} \label{eq:t2btgw}
    \tau^{(E)}_{GW}=S\times\tau^{(J)}_{2B},
\end{equation}
where $S\simeq 1.36$, as determined by numerically solving the Fokker-Planck equation \citep[see][]{Kaur_2024}.
Eq. (\ref{eq:t2btgw}) defines an effective loss-cone boundary in the $(r,r_p)$ plane, given by
\begin{equation}\label{eq:2BGW}
    \begin{aligned}
    \frac{r_{p, \rm lc}}{R_s}&=c_p\alpha^{2/5}\msbp{-1/5}\left(\frac{r}{R_h}\right)^{-1/2}\\
    &\simeq0.7\left(\frac{r}{R_h}\right)^{-1/2},    
    \end{aligned}
\end{equation}
where $\alpha=425\pi^2/\left(2048\sqrt{2}c_{7/4}\log\Lambda\right)\simeq0.11$ and $c_p=S^{2/5}=1.13$, as defined in \cite{Kaur_2024}. Above this line, i.e., for $r_p > r_{p,\rm lc}$, two-body scattering dominates the orbital evolution, leading to a random walk in $r_p$, while below it, the orbit shrinks and circularizes due to GW emission, as depicted in Fig. (\ref{fig:1}).
In the derivation of Eq. (\ref{eq:2BGW}), we assume that the sBHs follow a BW profile. 

Although Eq. (\ref{eq:2BGW}) was derived under the assumption that sBHs dominate the scattering, it remains valid in regions where the stars are the dominant scatterers. 
This is because the stars have a smaller cross section compared to the sBHs but they are more numerous. These two effects balance out when the BW profile of sBHs is normalized according to \cite{Linial_2022}, as discussed below (see Eqs. \ref{eq:RId} and \ref{eq:nBH}).

The combined effect of diffusion of sBHs in angular momentum due to two-body scattering and circularization due to GW emission may lead either to a slowly evolving EMRI or to a rapid plunge, where the sBH reaches a periapsis smaller than the mostly bound orbit, $r_p=4R_s$. 
The latter introduces a critical semimajor axis: \citep{Hopman_2005}
\begin{equation}\label{eq:Rc}
    \frac{R_c}{R_h}\simeq 0.03\left(\frac{\mstar/\mbh}{0.1}\right)^{-2/5},
\end{equation}
which is given by Eq. (\ref{eq:2BGW}) with $r_{p,\rm lc}(R_c)=4R_s$ (and see Fig. \ref{fig:1}). Thus, direct plunges originate from orbits with $r\gtrsim R_c$, while orbits with $r\lesssim R_c$ lead to the formation of EMRIs, as they enter the GW-dominated region and circularize before plunging into the SMBH.
Note that this dichotomy between the progenitors of EMRIs and plunges is less pronounced for $\Mbh\lesssim10^5 M_\odot$ \citep{Qunbar_2023}.

In the region where two-body scattering dominates the orbital evolution, the steady-state distributions of the sBHs and the stars satisfy a zero-flux solution \citep{BW_76,Linial_2022}.
Hence, at the outermost regions, up to $R_h$, the stars dominate the scattering and therefore settle into the BW profile and the sBHs number density follows a steeper profile, $\nbh(r)\propto r^{-3/2-\mbh/\left(4\mstar\right)}=r^{-4}$, as they efficiently migrate inward, following the result of \cite{Linial_2022}, who showed that $p_\bullet/p_\star \approx  \mbh/\mstar$. Since the sBH density increases at smaller distances, at a certain distance, defined below as $\RI$, the sBHs become the dominant scatterers. Therefore, for $r<\RI$, the sBHs follow the BW profile while the stars obtain a shallower profile, with $p_\star=\mstar/(4\mbh)$, and so $\nstar(r)\approxprop r^{-3/2}$. 
Note that the number of stars is normalized such that the total mass of the stars within the radius of influence is roughly twice the SMBH's mass \citep{Biney_Tremaine,Merritt_2004}, i.e., $\Nstar\left(R_h\right)\approx2\Mbh/\mstar$, while the normalization of the sBH distribution ensures a continuous zero-flux solution \citep{Linial_2022}:
\begin{equation} \label{eq:RId}
    \frac{\nbh\left(\RI\right)}{\nstar\left(\RI\right)}=\left(\frac{\mstar}{\mbh}\right)^{3/2}    
\end{equation}

The transition between the star-dominated scattering region to sBH-dominated one, $\RI$, as defined implicitly by Eq. (\ref{eq:RId}), is given by 
\begin{equation}\label{eq:RI}
    \frac{R_I}{R_h}=\fbh^{4/5}\msbp{-6/5}\simeq 0.06 \left(\frac{\fbh}{10^{-3}}\right)^{4/5},
\end{equation} 
where we define the sBH number fraction as
\begin{equation}\label{eq:fbh}
    \fbh=\frac{\Nbh\left(\RI\right)}{N_\star\left(R_h\right)} \,.
\end{equation}
Namely, the total number of sBHs within the sphere of influence, which is dominated by their number around $R_I$, $\Nbh(\RI)$, is a fraction $f_\bullet$ of the number of stars within the sphere of influence,  $N_\star\left(R_h\right)\approx2\Mbh/\mstar$.

Comparing $R_c$ (Eq. \ref{eq:Rc}) and $\RI$ (Eq. \ref{eq:RI}) introduces a critical sBH number fraction, 
\begin{equation}\label{eq:fbc}
    f^c_\bullet\simeq4.5\times10^{-4} \left(\frac{\mbh/\mstar}{10}\right),
\end{equation}
such that $\RI>R_c$ for $\fbh>f^c_\bullet$ and vice versa. 
The critical number fraction $f^c_\bullet$ distinguishes between two qualitatively different EMRI formation scenarios, as discussed in Section (\ref{sec:3}).

The GW-dominated region is characterized by a broken power law as well. At the immediate vicinity of the SMBH, up to a characteristic distance $\RII$, as defined below, the steady-state distribution is determined by the GW timescale (Eq. \ref{eq:tgw}). Hence, considering circular orbits, $\Nbh\left(r\right)\sim \tau^{(E)}_{GW}\propto r^4$. However, at greater distances, $r>\RII$, there is an effective replenishment of sBHs by two-body scattering, which leads to a shallower profile.
The profile due to the replenishment of circular orbits can be determined by equating the two-body-scattering-induced flux, in the $r_p$-direction at a given semimajor axis $\tilde{r}$, and the GW flux, in the $r$-direction:
\begin{equation} \label{eq:flxs}
    \frac{\Nbh(r)}{\tau^{(E)}_{GW}(r)}=\frac{\Nbh(\tilde{r})}{\tau^{(J)}_{2B}(\tilde{r})\log\Lambda_0},
\end{equation}
where $\log\Lambda_0=\log\left[r/r_{p,\rm lc}\left(r\right)\right]\approx\log\left(R_c/R_s\right)$, is roughly the size of the relevant loss-cone and $r=r_{p,\rm lc}(\tilde{r})$, as given by Eq. (\ref{eq:2BGW}).
In Eq. (\ref{eq:flxs}), we use the two-body timescale and GW timescale for nearly circular orbits, i.e., Eqs. (\ref{eq:t2b}) and (\ref{eq:tgw}) with $r\approx r_p$. Substituting the BW profile, $\nbh(\tilde{r})\propto\tilde{r}^{-7/4}$, yields $\Nbh\left(r\right)\propto r^{2}$.
Note that the two-body scattering flux assumes an empty loss-cone dynamics \citep{Lightman_1977}, as further discussed in Section \ref{sec:3}.

As mentioned above, $\RII$ is the characteristic distance that is supplemented by the flux of sBHs with initial semimajor axis of $\sim\min\left\{\RI,R_c\right\}$, namely $\RII=r_{p,\rm lc}\left(\min\left\{\RI,R_c\right\}\right)$ or
\begin{equation} \label{eq:RII}
    \frac{\RII}{R_s}=4\times\max\left\{1,\left(\frac{f_\bullet^c}{\fbh}\right)^{2/5}\right\}.
\end{equation}

Thus, the number density of sBHs is given by
    \begin{equation}\label{eq:nBH}
    \begin{aligned}
        \nbh&(r)=\frac{1}{2\pi R_h^3}\frac{\Mbh}{\mstar}\\
        \times&\left\{\def\arraystretch{1}\begin{tabular}{@{}l@{\quad}l@{}}
        $\fbh^{9/5}\left(\frac{\mstar}{\mbh}\right)^{-6/5}\left(\frac{r}{R_h}\right)^{-4}$ & $R_{I}<r<R_{h}$ \\
        $\left(\frac{\mstar}{\mbh}\right)^{3/2}\left(\frac{r}{R_h}\right)^{-7/4}$ & $\RGW<r<R_{I}$ \\
        $C_1\left(\frac{\mstar}{\mbh}\right)^{8/5}\left(\frac{R_s}{R_h}\right)^{-3/2}\left(\frac{r}{R_s}\right)^{-1}$ & $\RII<r<\RGW$ \\
        $C_2\fbh^{4/5}\left(\frac{\mstar}{\mbh}\right)^{4/5}\left(\frac{R_s}{R_h}\right)^{-3/2}\frac{r}{R_s}$ & $R_s\lesssim r<\RII$ \\ \end{tabular}\right.,
    \end{aligned}
    \end{equation}
where $R_{\rm GW}$ is determined such that the sBH distribution is continuous, qualitatively corresponding to the distance where $\tau^{(E)}_{GW}(r)\sim\tau^{(J)}_{2B}(r)$, 
\begin{equation}\label{eq:RGW}
\begin{aligned}
    \frac{\RGW}{R_s}&=\left(\frac{\log\Lambda_0\alpha^{1/5}}{c_p^2}\right)^{4/3}\msbp{-2/15}\left(\frac{R_s}{R_h}\right)^{-1/3}\\
    &\approx 2\times10^3\left(\frac{\Mbh}{\Mmw}\right)^{-1/6},
\end{aligned}
\end{equation}
and the numerical order unity coefficients $C_1$ and $C_2$ are given by
\begin{equation}
    \begin{aligned}
        C_1&=\frac{c_p^2}{\alpha^{1/5}\log\Lambda_0}\approx0.2,\\    
        C_2&= \frac{\min\left\{1,f_\bullet^c/\fbh\right\}^{4/5}}{\alpha\log\Lambda_0}\\
        &\approx0.9\min\left\{1,f_\bullet^c/\fbh\right\}^{4/5},
    \end{aligned}
\end{equation}
where we have substituted $\log\Lambda=\log\Lambda_0=10$.

For completeness, we present the corresponding distribution of the stars, from $R_h$ to the tidal radius, where the SMBH tidal force is comparable to the star's self-gravity
\begin{equation} \label{eq:Rt}
    \frac{R_t}{R_s}\approx \frac{R_\star}{R_s}\left(\frac{\Mbh}{\mstar}\right)^{1/3}\simeq 9\left(\frac{\Mbh}{\Mmw}\right)^{-2/3},
\end{equation}
where $R_{\star}=R_\odot$ is the radius of the star.
However, unlike the sBH distribution, collisions are expected to significantly modify the star distribution \citep{Sari_Fragione_2019,Rose_2023,Balberg_2023}, hence we present only a qualitative analysis and leave a detailed calculation for a future work.

\begin{figure}[ht!] 
    \centering
    \includegraphics[width=8.6cm]{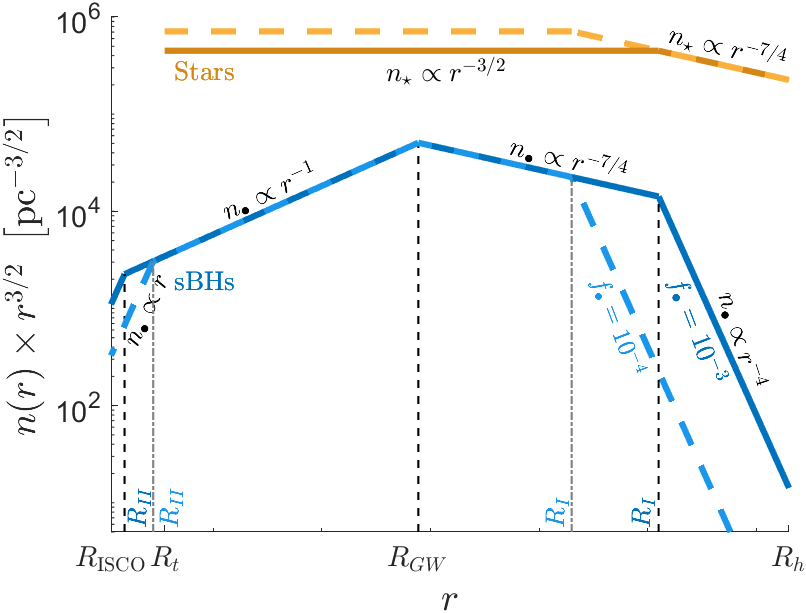}
    \caption{Steady-state distributions of the sBHs (blue lines) and the stars (yellow lines), as given by Eqs. (\ref{eq:nBH}) and (\ref{eq:nstar}). We consider a SMBH with $\Mbh=4\times10^6 M_\odot$ and sBH number fractions $\fbh=10^{-3}$ (solid lines) and $\fbh=10^{-4}$ (dashed lines), as defined in Eq. (\ref{eq:fbh}). We plot $n(r)\times r^{3/2}$, which corresponds to the commonly used phase-space density, $f(E)$. The star distribution extends from $R_h$ to the tidal radius, $R_t$ (Eq. \ref{eq:Rt}), while the sBHs can reach $R_{\rm ISCO}=3R_s$.}
     \label{fig:2}
\end{figure}

As mentioned above, at the outer parts of the NSC, where two-body scatterings dominate the orbital evolution, the stars distribution can be described by a broken power law, with $\nstar\left(r\geq\RI\right)\propto r^{-7/4}$ and $\nstar(r<\RI)\approxprop r^{-3/2}$. Considering only two-body scattering and GW emission, the stars qualitatively follow the flatter profile, $\nstar\left(r\right)\approxprop r^{-3/2}$, all the way\footnote{The stars at $r\lesssim\RGW$ are populated by two-body scattering replenishment via highly eccentric orbits. However, in practice, they follow almost the same power law, as it changes from $p_\star=\mstar/4\mbh$ to $p_\star=-\mstar/2\mbh$. Therefore, we neglect this transition and assume a constant $p_\star\approx0$ at this region.} to $r\sim\max\left\{\RII,R_t\right\}$.
Therefore, the number density of the stars, assuming $R_t>\RII$ (as depicted in Fig. \ref{fig:2}), is roughly   
\begin{equation}\label{eq:nstar}
\begin{aligned}
        n&_\star(r)\sim\frac{1}{2\pi R_h^3}\frac{\Mbh}{\mstar}\\
        &\times\left\{\def\arraystretch{1}\begin{tabular}{@{}l@{\quad}l@{}}
        $\left(\frac{r}{R_h}\right)^{-7/4}$ & $R_{I}<r<R_{h}$ \\
        $\fbh^{-1/5}\left(\frac{\mstar}{\mbh}\right)^{3/10}\left(\frac{r}{R_h}\right)^{-3/2}$ & $R_t<r<\RI$ \\
\end{tabular}\right..
\end{aligned}
\end{equation}
If $\RII>R_t$, which corresponds to $\fbh\lesssim10^{-4}$, the replenishment of self-scattered stars would dominate the feeding rate of stars to the orbits below $\RII$ and would therefore follow $\nstar\propto r^{-1}$, equivalently to the sBH distribution for $\RII<r<\RGW$. 

In Fig. (\ref{fig:2}), we present the density distributions of the stars and sBHs (Eqs. \ref{eq:nBH} and \ref{eq:nstar}), respectively. 
Note that the intermediate region of the sBH distribution, $\nbh\left(\RII<r<\RI\right)$, and the outer part of the star distribution, $\nstar\left(\RI<r<R_h\right)$, are universal, in the sense that their normalization does not depend on the sBH number fraction $\fbh$. We demonstrate that in Fig. (\ref{fig:2}), which depicts the distributions for $\fbh=10^{-3}$ (solid lines) and $\fbh=10^{-4}$ (dashed lines).

Our analysis assumes that the NSC reaches a steady state, which is valid for clusters surrounding SMBHs with masses $\Mbh\lesssim10^7 M_\odot$, for which the relaxation time is shorter than the Hubble time \citep{Biney_Tremaine,Bar_Or_2013}. 

\section{EMRI characteristics: formation rate and eccentricity distribution} \label{sec:3}

In a given galaxy, the instantaneous distribution of sBHs is described by Eq. (\ref{eq:nBH}), ensuring that the flux induced by GW emission aligns with the corresponding two-body scattering flux, which sets the rate at which sBHs enter the GW-dominated region (see Eq. \ref{eq:flxs}).
Consequently, the EMRI formation rate is given by \citep{Hopman_2005}:
\begin{equation}\label{eq:rateE}
    \begin{aligned}
        \Gamma_{\rm EMRI}&=\int_{\RGW}^{R_c} dr\frac{4\pi r^2\nbh\left(r\right)}{\tau^{(J)}_{2B}(r)\log\Lambda_0}\\
        & \approx 260\left(\frac{\Mbh}{M_{\rm MW}}\right)^{-1/4}y_E(\fbh)\ {\rm Gyr^{-1}},
    \end{aligned}
\end{equation}
with $y_E(\fbh)\approx\left\{\def\arraystretch{0.8}\begin{tabular}{@{}l@{\quad}l@{}}
  $1$ & $\fbh> f^c_\bullet$ \\
$850\fbh^{4/5}$ & $\fbh< f^c_\bullet$
\end{tabular}\right.$.

Note that, generally, the $\log\Lambda_0$ term introduces a logarithmic dependence on the semimajor axis, which we simplify by taking it as a constant. We assume $\log\Lambda_0=10$, which roughly corresponds to its value at $\min\left\{\RI,R_c\right\}$, where most of the EMRIs are originated from.

The EMRI formation rate is dominated by the flux of sBHs from $\min\left\{\RI,R_c\right\}$, since $\Gamma_{\rm EMRI}\propto r$, for $r<\RI$ and $\Gamma_{\rm EMRI}\propto r^{-5/4}$, for $r>\RI$. Moreover, the value of the maximal EMRI formation rate, achieved for $\fbh>f_\bullet^c$, does not depend on $\fbh$. 

Direct plunges - namely, sBHs with semimajor axis $r\gtrsim R_c$ that may reach $r_p\lesssim 4R_s$ by two-body scattering, before circularizing due to GW emission - occur at a rate
\begin{equation}\label{eq:rateP}
\begin{aligned}
    \Gamma_{\rm Plunge}&=\int_{R_c}^{R_h} dr\frac{4\pi r^2\nbh\left(r\right)}{\tau^{(J)}_{2B}(r)\log\Lambda_0}\\    
    & \approx 210\left(\frac{\Mbh}{M_{\rm MW}}\right)^{-1/4}y_P(\fbh)\ {\rm Gyr^{-1}},
\end{aligned} 
\end{equation}
where $y_P(\fbh)\approx\left\{\def\arraystretch{0.75}\begin{tabular}{@{}l@{\quad}l@{}}
    $1000\fbh^{4/5}$ & $\fbh> f^c_\bullet$ \\
  $\left(\fbh/f^c_\bullet\right)^{9/5}$ & $\fbh< f^c_\bullet$
\end{tabular}\right.$.

In Table (\ref{tab:1}), we present the EMRI and plunge formation rates for different values of $\fbh$. Note that the formation rates, both of EMRIs and plunges, are proportional to the orbital period at the radius of influence, $\Gamma_{\rm EMRI,Plunge}\propto 1/P(R_h)\propto M_{\rm BH}^{3/\beta-1}$, as previously derived by \cite{Hopman_2005}. Additionally, we assume $\RGW\ll\RI,R_c\ll R_h$, and therefore neglect the contributions from $\RGW$ and $R_h$ in Eqs. (\ref{eq:rateE}) and (\ref{eq:rateP}), respectively. 

The above calculation assumes an empty loss-cone dynamics \citep{Lightman_1977}, which is valid for the relevant SMBH masses. A transition to a full loss-cone region, where $P(r)/\tau^{(J)}_{2B}\left(r,r_{p,\rm lc}\left(r\right)\right)\gg\log\Lambda_0$ \citep{Vasiliev_2013},  corresponds to $\Mbh\ll10^4 \Msun (f_{\bullet}/10^{-3})^{2/3}$ and therefore is not considered here.
\begin{table}[ht!]
    \centering
    \begin{tabular}{ |c|c|c| } 
    \hline
    $\fbh$ & $\Gamma_{\rm EMRI}\ \left[{\rm Gyr}^{-1}\right]$ & $\Gamma_{\rm Plunge}\ \left[{\rm Gyr}^{-1}\right]$ \\ 
    \hline
    $10^{-2}$ & $260$ & $5000$ \\ 
    $10^{-3}$ & $260$ & $600$ \\ 
    $10^{-4}$ & $130$ & $15$ \\ 
    $10^{-5}$ & $20$ & $0.22$ \\ 
    \hline
    \end{tabular}
    \caption{Formation rates of EMRIs (Eq. \ref{eq:rateE}) and plunges (Eq. \ref{eq:rateP}) for different values of the sBHs number fraction $\fbh$, assuming an SMBH with $\Mbh=4\times10^6M_\odot$ and a population of sBHs with mass $\mbh=10M_\odot$.}
    \label{tab:1}
\end{table}

\begin{figure}[!ht] 
    \centering
    \includegraphics[width=8.6 cm]{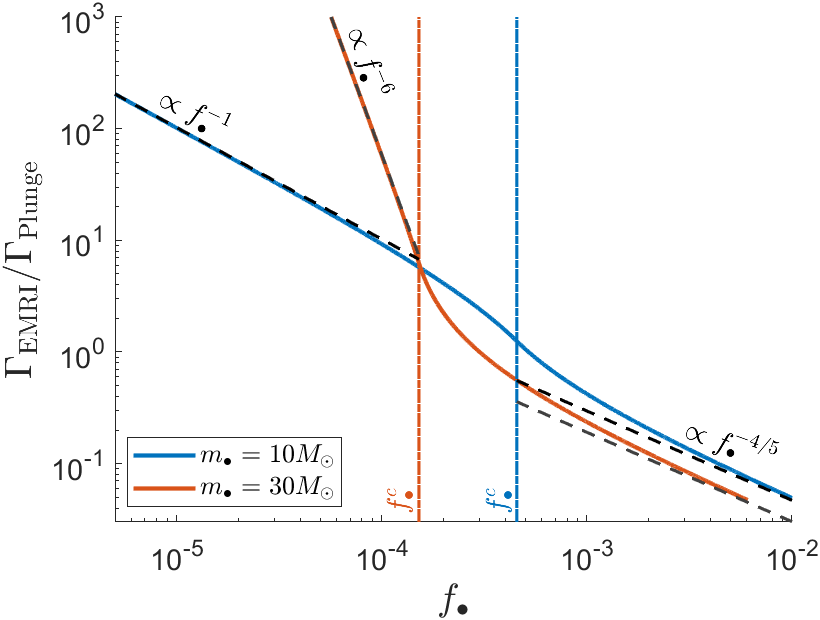}
    \caption{EMRI-to-plunge ratio as a function of the sBH number fraction, assuming sBHs of mass $\mbh=10M_\odot$ (blue line) and $\mbh=30M_\odot$ (red line), as calculated from Eqs. (\ref{eq:rateE}) and (\ref{eq:rateP}). The dashed-dotted lines mark the critical number density, $f_\bullet^c$ (Eq. \ref{eq:fbc}), for each sBH mass, with respective colors, and the dashed lines present the ratio in the limiting cases of scarce ($\fbh\ll f_\bullet^c$) and abundant ($\fbh\gg f_\bullet^c$) sBH population (Eqs. \ref{eq:E2P1} and \ref{eq:E2P2}), respectively.}
    \label{fig:3}
\end{figure}

In Fig. (\ref{fig:3}) we present the EMRI-to-plunge ratio, as given in Eqs. (\ref{eq:rateE}) and (\ref{eq:rateP}).
Considering the two limiting cases, if the sBHs are scarce,
\begin{equation}\label{eq:E2P1}
    \left.\frac{\Gamma_{\rm EMRI}}{\Gamma_{\rm Plunge}}\right|_{\fbh\ll f^c_\bullet}\approx 100\left(\frac{10^{-5}}{\fbh}\right).
\end{equation}
while if they are abundant,
\begin{equation}\label{eq:E2P2}
    \left.\frac{\Gamma_{\rm EMRI}}{\Gamma_{\rm Plunge}}\right|_{\fbh\gg f^c_\bullet}\approx 
    0.05\left(\frac{10^{-2}}{\fbh}\right)^{4/5}.
\end{equation}
For the full expressions and the dependence on the sBH mass, see Appendix \ref{app:a}.

Our estimated EMRI formation rate for a Milky Way-like galaxy is comparable to previous results in the literature \citep[e.g.,][]{Hopman_2006,Amaro_Seoane_2011,Merritt_2015,Aharon_2016,Aceves_2022,Broggi_2022}. However, we predict fewer plunges, compared to earlier studies, since in our model the sBHs are concentrated around $\RI\ll R_h$ rather than around $R_h$, as in the single-power-law models \citep[e.g.,][]{Bar_or_2016}. Notably, for a low sBH number fraction, $\fbh\lesssim 5\times10^{-3}$, the EMRI rate surpasses the plunge rate. see Section \ref{sec:comp} for further discussion.

Finally, we note that EMRIs and plunges may originate from sBH binary disruptions \citep{Hills_1988} as well. The contribution of such a scenario depends linearly on the sBH binary fraction and tends to increase more prominently the EMRI formation rate for a significantly large binary fraction \citep[see][for further details]{Sari_Fragione_2019}. We ignore this effect here.

\subsection{EMRI Eccentricity Distribution}
The eccentricity evolution of the sBH orbits at the GW-dominated region, from an initial periapsis $r_{p,i}$ with $e_i\approx1$ up to the mostly bound orbit, can be evaluated using \cite{Peters_1964}, such that the eccentricity $e$ satisfies
\begin{equation}\label{eq:e_ev}
    g(e)=\frac{4R_s}{r_{p,i}}=\left(\frac{r_i}{R_c}\right)^{-1/2},
\end{equation}
where
\begin{equation}\label{eq:ge}
    g(e)=\frac{2e^{12/19}}{(1+e)}\left(\frac{304+121e^2}{425}\right)^{870/2299}.
\end{equation}
Note that in the second equality in Eq. (\ref{eq:e_ev}), we assume that $r_{p,i}$ is along the loss-cone boundary, hence it can be related to a semimajor axis $r_i$ using Eq. (\ref{eq:2BGW}), together with the definition of $R_c$ (Eq. \ref{eq:Rc}).

Thus, the minimal eccentricity is given by $g(e_{\min})\simeq\frac{4Rs}{\RGW}$, or
\begin{equation}\label{emin}
e_{\min}\approx10^{-5}\left(\frac{\Mbh}{\Mmw}\right)^{19/72},
\end{equation}
where we used $g(e\ll1)\propto e^{12/19}$. On the other hand, the maximal eccentricity stems from sBHs with $r_i\approx R_c$, and therefore
\begin{equation}\label{eq:emax}
    (1-e_{\max})\approx\frac{4Rs}{Rc}\approx10^{-5}\left(\frac{\Mbh}{\Mmw}\right)^{1/2}.
\end{equation}

The eccentricity distribution depends on the number fraction of the sBHs $\fbh$. If $\fbh>f^c_\bullet$, i.e., $\RI>R_c$, all the EMRI progenitors initially follow a BW profile, leading to a two-body-scattering-induced flux $\mathcal{F}_{\rm 2B}\propto\Gamma_{\rm EMRI}\propto r_i$ (as evident from Eq. \ref{eq:rateE}) and, hence, using Eq. (\ref{eq:e_ev}): 
\begin{equation}\label{eq:pe1}
    \left.p(e)\right|_{\fbh\geq f^c_\bullet}=2g(e)g'(e),
\end{equation}
as derived by \cite{Linial_2023} in the context of stellar EMRIs. This distribution peaks at $e\simeq0.08$, and scales as $p(e\ll1)\propto e^{5/19}$.

\begin{figure}[!ht] 
    \centering
    \includegraphics[width=8.6 cm]{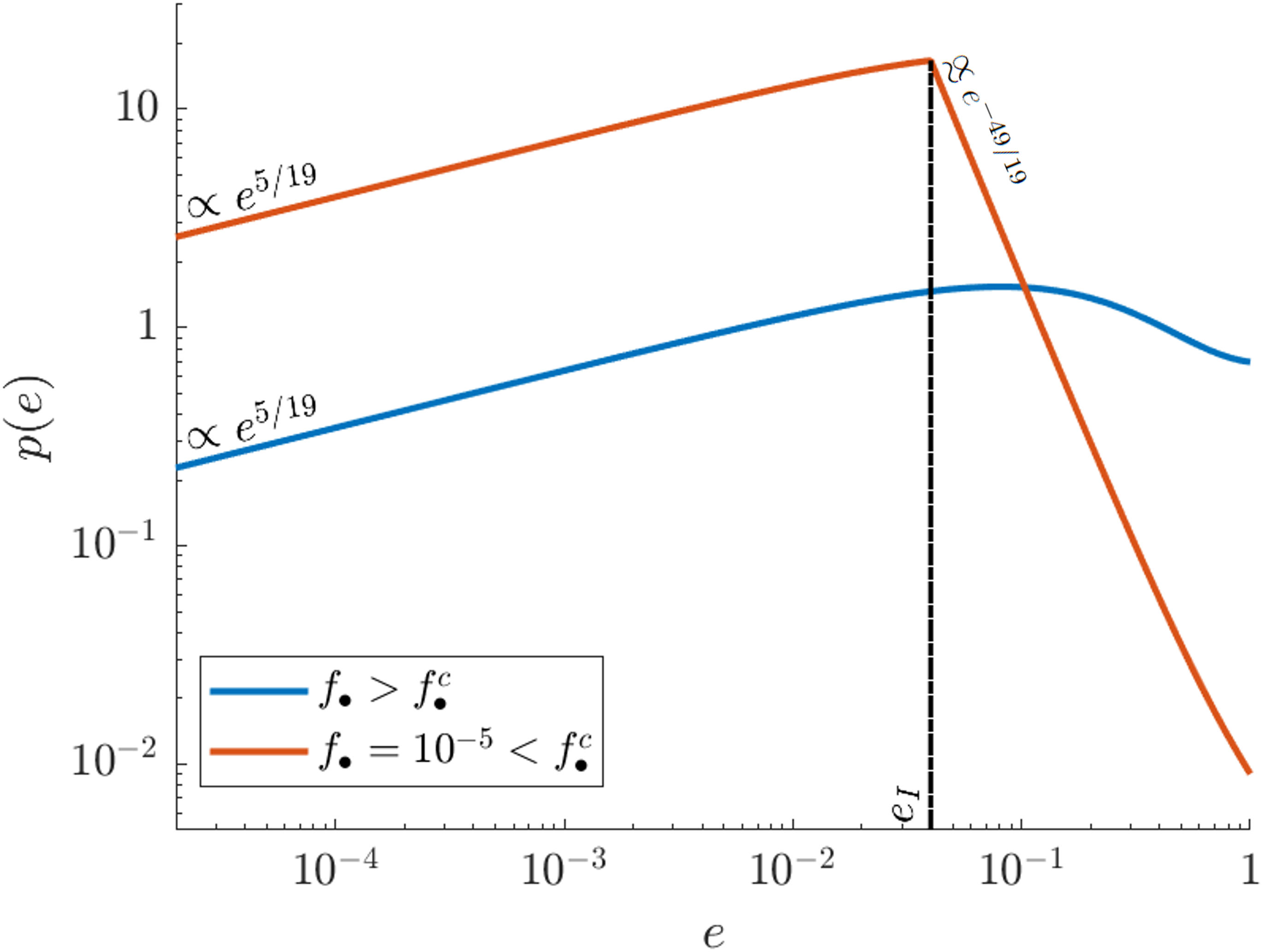}
    \caption{EMRI eccentricity probability density function at the mostly bound orbit. The blue line corresponds to the case where the sBH number fraction is larger than the critical value, $\fbh\geq f_\bullet^c$, where the distribution is independent of the specific value of $\fbh$ and peaks at $e \simeq 0.08$. The red line exhibits an example of the complementary case, $\fbh=10^{-5}<f_\bullet^c$, where the distribution peaks at $e_I\simeq0.04$ (as given by Eq. \ref{eq:geI}) and falls steeply, as $p(e\gtrsim e_I)\approxprop e^{-49/19}$. In both cases, the distributions satisfy $p(e\ll1)\propto e^{5/19}$.}
    \label{fig:4}
\end{figure}

In the second case, where $\fbh<f^c_\bullet$, there is a characteristic eccentricity, $e_I$, given by the implicit relation
\begin{equation} \label{eq:geI}
    g(e_I)\approx\frac{4R_s}{\RII},
\end{equation}
yielding, for example, $e_I\approx0.2$ or $e_I\approx0.04$, for $\fbh=10^{-4}$ and $\fbh=10^{-5}$, respectively.
For $e<e_I$ we get as before $p(e)\propto g(e)g'(e)$, while the EMRIs with $e>e_I$ originate from $r_i>\RI$, and hence follow a different two-body-scattering-induced flux, $\widetilde{\mathcal{F}}_{2B}\propto r_i^{-5/4}$, which leads to $p(e)\propto g(e)^{-7/2}g'(e)$. Taking into account the probability distribution normalization, we get 
\begin{equation}\label{eq:pe2}
    \begin{aligned}
\left.p(e)\right|_{\fbh<f^c_\bullet}\simeq&\frac{10g'(e)g(e)}{9g(e_I)^{-5/2}-4}\\
    &\times\left\{\def\arraystretch{1}\begin{tabular}{@{}l@{\quad}l@{}}
        $g(e_I)^{-7/2}$ & $e_{\min}\leq e\leq e_I$ \\
        $g(e)^{-9/2}$ & $e_I<e\leq e_{\max}$ \\
        \end{tabular}\right. ,    
    \end{aligned}
\end{equation}
which reproduces Eq. (\ref{eq:pe1}) in the limit $e_I\rightarrow1$. Additionally, the distribution scales as $p(e\ll1)\propto e^{5/19}$, while $p(e\gtrsim e_I)\approxprop e^{-49/19}$. The eccentricity distributions, Eqs. (\ref{eq:pe1}) and (\ref{eq:pe2}), are presented in Fig. (\ref{fig:4}).

\section{EMRIs signature as GW Sources} \label{sec:4}
Motivated by the tendency toward low eccentricities in the vicinity of the SMBH, as evident from Fig. (\ref{fig:4}), we focus on EMRIs where the sBH follows an adiabatic, quai-circular inspiral - namely it slowly descends from one circular orbit to another (for further details, see Appendix \ref{app:c}). 
In this case, the GWs are emitted predominantly at twice the orbital frequency:
\begin{equation}\label{eq:st1}
    f=\frac{1}{\pi}\sqrt{\frac{G\Mbh}{r^3}}.
\end{equation}

The inclination-averaged characteristic strain of the GWs is given by \citep{Finn_2000}:
\begin{equation}\label{eq:hc}
            h_c=\varepsilon_{h}\frac{8}{\sqrt{5}}\left(\pi \fobs\right)^{2/3}\frac{\left(G\Mcobs\right)^{5/3}}{c^4d_L(z)}\sqrt{\fobs T^{\rm obs}_{i}},
\end{equation}
where $\mathcal{M}=\mbh^{3/5}\Mbh^{2/5}$ is the chirp mass, $\Mcobs=\mathcal{M}\left(1+z\right)$ is the observed (redshifted) chirp mass, $\fobs=f/(1+z)$ is the observed GW frequency, $\varepsilon_{h}$ is the general relativity (GR) correction term (see Appendix \ref{app:c}), and $d_L(z)$ is the luminosity distance (see Appendix \ref{app:b}). 
The characteristic strain increases as the square root of the number of orbits the sBH undergoes at the frequency $\sim\fobs$, therefore
\begin{equation} \label{eq:Ti}
    T^{\rm obs}_i=\min\left\{\TL,\frac{\fobs}{\dot{f}_{\rm obs}}\right\},    
\end{equation}
where $\TL=4\ \rm yr$ is the LISA mission time and
\begin{equation}\label{eq:st3}
    \frac{\fobs}{\dot{f}_{\rm obs}}=\varepsilon_f\frac{5}{96}\left(\pi\fobs\right)^{-8/3}\left(\frac{G\Mcobs}{c^3}\right)^{-5/3},
\end{equation}
where the dot corresponds to a derivative with respect to time in the detector frame and $\varepsilon_f$ is the GR correction (see Appendix \ref{app:c}). 
Qualitatively, $T^{\rm obs}_i$ separates between static, monochromatic sources ($\fobs/\dot{f}_{\rm obs}>\TL$) and evolving sources ($\fobs/\dot{f}_{\rm obs}<\TL$).

The number of sources, per frequency, SMBH mass, and redshift, is given by:
\begin{equation} \label{eq:nSources}
\begin{aligned}
    \mathcal{N}&=\frac{\dd N}{\dd\left(\log\fobs\right)\dd\left(\log\Mobs\right)\dd z}\\
    &=\frac{2}{3}\Nbh\frac{\TL}{T^{\rm obs}_i}\frac{\dd \rho_{M}}{\dd\left(\log\Mobs\right)}\frac{\dd V_c}{\dd z},
\end{aligned}
\end{equation}
where $\Nbh(M,f)$ is the sBH number in a given galaxy with orbital frequency $f/2$ (as given by Eqs. \ref{eq:Nn}, \ref{eq:nBH}, and \ref{eq:st1}), $\Mobs=\Mbh(1+z)$ is the observed SMBH mass, $V_{c}$ is the comoving volume (see Appendix \ref{app:b}), and the SMBH mass function is given by \citep{Babak_17}:
\begin{equation}\label{eq:rhoSMBH}
    \frac{\dd{\rho}_{M}}{\dd\left(\log\Mobs\right)} = \rho_0 \left( \frac{\Mbh}{\Mmw}\right)^{-\zeta},
\end{equation}
with $\rho_0 = 0.005 \, {\rm Mpc}^{-3} $ and $\zeta = 0.3$.
Note that in Eq. (\ref{eq:nSources}), the factor of $\TL/T^{\rm obs}_i$ accounts for evolving sources that sweep a range of frequencies and the $2/3$ prefactor stems from taking the derivative with respect to the frequency rather than the semimajor axis.

For evolving sources, $\Nbh\TL/T_i\sim \Gamma_{\rm EMRI}\times\TL$ - namely, the EMRI formation rate times the LISA mission time. For steady sources, the probability of having a source at a given galaxy corresponds to the number of sBHs with the relevant semimajor axis.

The total number of sources, per logarithmic frequency bin, with characteristic strain larger than $h_c$, is given by integrating Eq. (\ref{eq:nSources}) with respect to $\Mobs$ and $z$, which is related to $h_c$ by Eq. (\ref{eq:hc}).

\begin{figure}[!ht] 
    \centering
    \includegraphics[width=8.6 cm]{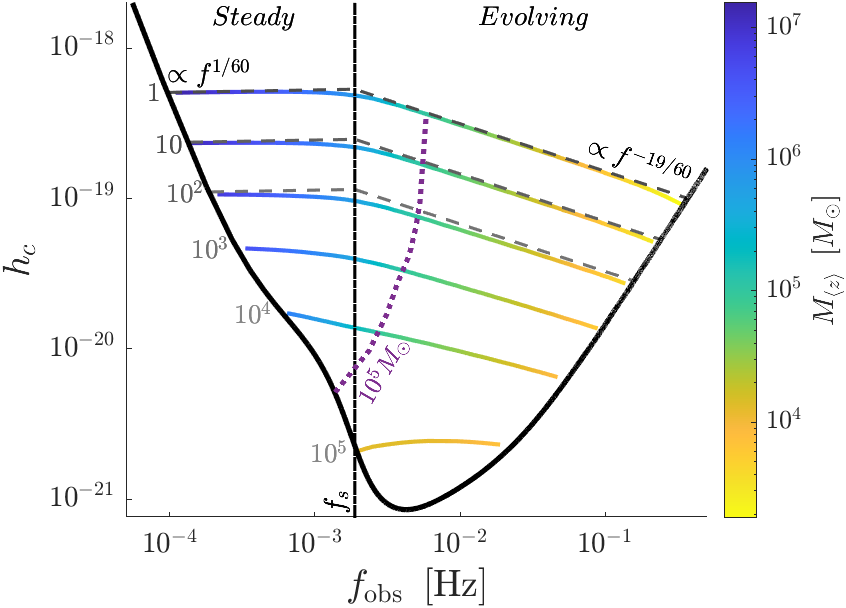}
    \caption{Characteristic strain of EMRIs as a function of the observed frequency. The black line is the LISA sensitivity curve \citep{Robson_2019} and the colored lines present contours of the observed sources numbers, between $1$ and $10^5$, in a logarithmic frequency bin. The dashed lines present an analytical estimation at $z\ll1$ (Eq. \ref{eq:f3}), scaling as $f^{1/60}$ at low frequencies and $f^{-19/60}$ at high frequencies. The color bar presents the source-frame SMBH mass that corresponds to the average redshift of the observed sources. The purple dotted line depicts the average SMBH mass $M_{\langle z\rangle}=10^5M_\odot$.}
    \label{fig:5}
\end{figure}

\begin{figure}[!ht] 
    \centering
    \includegraphics[width=8.6cm]{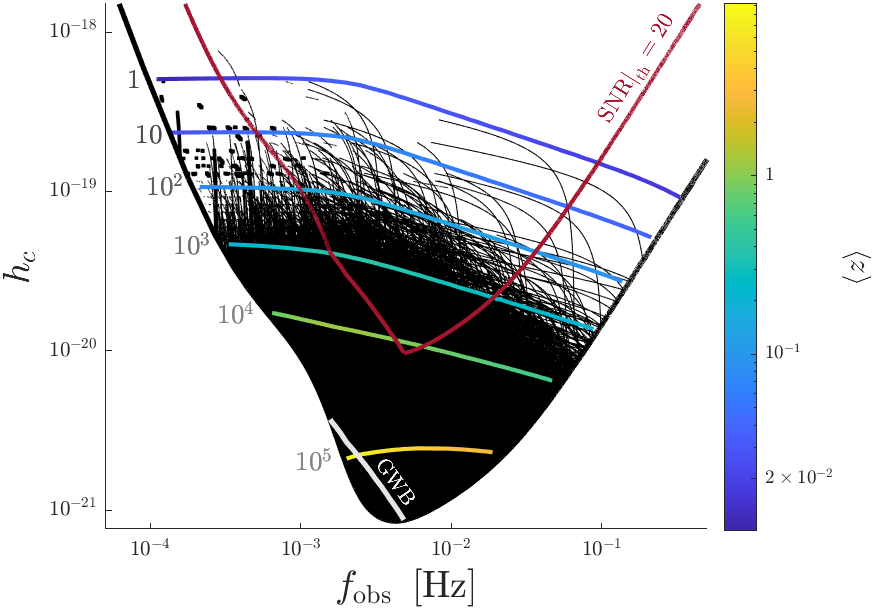}
        \caption{A realization of the observed EMRIs; each black line represents an EMRI, chosen according to the distribution given in Eq. (\ref{eq:nSources}), assuming an observation time $\TL=4\ \rm yr$. As in Figure \ref{fig:5}, the black thick line is the LISA sensitivity curve and the colored lines present contours of the observed sources numbers. The color bar presents the average redshift of the observed sources. The white line corresponds to the GWB (Eq. \ref{eq:hcb}), assuming a limit of one source per $4$ frequency bins, $\delta f=1/\TL$. Roughly $2\times10^3$ resolvable sources lie above the red line, which depicts the $\left.\rm {SNR}\right|_{\rm th}=20$ threshold.}
    \label{fig:6}
\end{figure}

We present in Fig. (\ref{fig:5}) the expected number of sources, assuming an sBH number fraction $\fbh=10^{-3}$, along with the LISA sensitivity curve, $h_{c,\rm {LISA}}$ \citep[as given by][]{Robson_2019}. 
 In Fig. (\ref{fig:6}), we present a realization of our predicted sources distribution (Eq. \ref{eq:nSources}), where each black line represents an EMRI.
 Thus, for example, we expect that during the LISA mission time, roughly one EMRI will be observed, per logarithmic frequency bin, with characteristic strain greater than the $\dd N/\dd\left(\log\fobs\right)=1$ line in Figs. (\ref{fig:5}) and (\ref{fig:6}). 

In addition, we estimate the redshift expected value, $\langle z\rangle$, and its corresponding source-frame SMBH mass, $M_{\langle z\rangle}$, at a given characteristic strain and observed frequency.  We present the values of $M_{\langle z \rangle}$ and $\langle z \rangle$ along the contours using color bars in Figs. (\ref{fig:5}) and (\ref{fig:6}), respectively. In Fig. (\ref{fig:5}), we present with a purple dotted line the contour of $M_{\langle z\rangle}=10^5 M_\odot$; below it, lighter SMBHs dominate the distribution, where the SMBH mass function and the $\Mbh-\sigma_h$ relation are less constrained.  

\subsection{Analytical estimation of the observed source distribution}\label{sec:An}
We can integrate the source distribution (Eq. \ref{eq:nSources}), analytically by: (a) considering the local Universe, i.e., sources at $z\ll1$, where the geometry is roughly flat; and (b) approximating the GR corrections as an effective numerical prefactor $\varepsilon$ (as discussed in Section \ref{app:c}). 

Under these assumptions, substituting Eqs. (\ref{eq:st1}) and (\ref{eq:hc}) yields
\begin{equation}\label{eq:f2}
\begin{aligned}
\left.\frac{\dd N}{\dd\log f}\right|_{h_c}=&\frac{4096\pi^3}{45\sqrt{5}}\frac{G^5\mbh^3f^{7/2}\TL\rho_0\varepsilon^3}{c^{12}h_c^3}\\
&\times \int \Nbh\Mbh^{1-\zeta} T_i^{1/2}\dd\Mbh.   
\end{aligned}
\end{equation}

The number of sources at each frequency bin is dominated by the SMBH with mass
\begin{equation}\label{eq:Mmax}
        M_{\max}\approx 0.8\Mmw \left(\frac{1 \rm mHz}{f}\right).
\end{equation}
This can be understood as the SMBH for which the orbital frequency at its ISCO corresponds to the observed GW frequency, reduced by a numerical prefactor that encapsulates the shortened merger time from the ISCO \citep{Ori_2000,Buonanno_2000,Finn_2000}.
As discussed in Appendix \ref{app:c}, this prefactor, which we take as $0.8$ in Eq.  (\ref{eq:Mmax}), ranges between $\approx0.65$ for evolving sources (at high frequencies, $f\gtrsim f_s$, as defined in Eq. \ref{eq:fs}), to $\approx1$ for steady sources (at low frequencies, $f\lesssim f_s$).

We further introduce a characteristic frequency, $f_s$, for which the merger time from the SMBH's ISCO is roughly the LISA mission time:
\begin{equation}\label{eq:fs}
    f_s\approx\frac{1}{2\pi}\sqrt{\frac{5c^3}{4G\mbh\TL}}\approx 2\rm mHz,
\end{equation}
Thus, at low frequencies, $f\lesssim f_s$, nearly steady, monochromatic EMRIs are the dominant sources, while at higher frequencies, $f\gtrsim f_s$, evolving sources, which sweep across a significant portion of the LISA band, become more prominent.

Finally, we get from Eq. (\ref{eq:f2}), using Eqs. (\ref{eq:nBH}) and (\ref{eq:Ti}), the characteristic strain as a function of the observed frequency, for a given number of sources:
\begin{equation} \label{eq:f3}
\begin{aligned}
    h_c\approx& \, 5\times10^{-19}\left(\frac{f}{f_s}\right)^{\delta}\left(\frac{\dd N}{\dd\log f}\right)^{-1/3}\\
    &\times \left(\frac{\fbh}{f_\bullet^c}\right)^{4/15}\frac{\xi\left(\min\left\{\fbh,f_\bullet^c\right\}\right)}{\xi\left(f_\bullet^c\right)}.
\end{aligned}
\end{equation}
For steady sources $(f\lesssim f_s)$, $\delta=1/60$ and $\xi\left(x\right)\simeq \left(1-17x^{47/100}\right)^{1/3}$.
For evolving sources $(f\gtrsim f_s)$, $\delta=-19/60$ and $\xi\left(x\right)\simeq \left(1-5x^{27/100}\right)^{1/3}$. 

The analytical estimation agrees with the full numerical results at relatively large characteristic strains, $h_c\gtrsim10^{-19}$, as presented in Fig (\ref{fig:5}), corresponding to nearby sources, and deviates when considering cosmological distances, which contribute to the lower values of $h_c$.

\subsection{GW Background}\label{sec:GWB}
In addition to the resolvable sources, with signal-to-noise ratio (SNR) greater than a given detectability threshold, $\left.\rm {SNR}\right|_{\rm th}$, there is an accumulation of a stochastic GWB, induced by a population of unresolved sources. This effectively produces a background noise, a so-called confusion noise, which may decrease LISA's sensitivity, if it exceeds the intrinsic instrumental noise - as in the case of the galactic white dwarf binaries \citep{Nelemans_2001,Timpano_2006,Ruiter_2010,Nissanke_2012,Cornish_2017}.   

The noise introduced by the EMRI GWB can be estimated as the characteristic strain for which there is one source per few frequency bins, $a\times\delta f$, with $\delta f=1/\TL$. We take $a=3$ as an effective parameter corresponding to the signal smearing, due to LISA's motion, as well as the limitations of the data analysis procedure \citep{Hughes_2002,Barack_2004,Timpano_2006}.    
A numerical fit to the GWB yields
\begin{equation}\label{eq:hcb}
    h_{c,\rm GWB}\simeq2.1\times10^{-21}\left(\frac{f}{2.5\ \rm mHz}\right)^{-1.26}.
\end{equation}

In Fig. (\ref{fig:6}), we present the GWB in the range $\sim 1-5\ \rm mHz$, where it surpasses the LISA instrumental noise, by up to a factor of $\approx2$ around $2.5\ \rm mHz$. 

Given the EMRI GWB, we can estimate the number of resolvable sources.
The SNR of a source, averaged over sky location, inclination, and polarization, is given by \citep{Robson_2019}:
\begin{equation}\label{eq:SNR}
{\rm SNR}^2=\frac{16}{5}\int \dd\log \fobs\left(\frac{h_c}{h_{c,{\rm noise}}}\right)^2,    
\end{equation}
where $h_{c,{\rm noise}}=\max\left\{h_{c,\rm LISA},h_{c,\rm GWB}\right\}$.
We estimate the number of resolvable sources by averaging the number of sources with ${\rm SNR}\geq\left.\rm {SNR}\right|_{\rm th}$, calculated using Eq. (\ref{eq:SNR}), over 10 different realizations of our source distribution (Eq. \ref{eq:nSources}); one of these realizations is presented in Fig. (\ref{fig:6}). 

Considering a 4 yr mission time, we expect about $2\times10^3$ resolvable sources, with SNR above $\left.\rm {SNR}\right|_{\rm th}=20$. 
A lower SNR threshold, e.g., $\left.\rm {SNR}\right|_{\rm th}=8$, increases the number of resolvable sources by a factor of $\approx5$. 
Note that other sources and formation channels, e.g., EMRIs from SMBH binaries \citep{Naoz_2023}, may induce a stronger GWB, which will reduce the number of resolvable sources estimated here.

\section{Comparison with Previous Works} \label{sec:comp}
In this work, we present several modifications to the widely used steady-state distributions in NSCs and the EMRI formation mechanism. The following section summarizes the main differences between our analysis and previous results in the literature.

On the outskirts of the sphere of influence, external to $R_I$, where strong segregation occurs, we apply the result of \cite{Linial_2022} for the steady-state distribution. Therefore, our sBH density profile, $\nbh\propto r^{-4}$, is steeper than the ``strong mass segregation" profile of \cite{AH_09}.
Moreover, the distributions of the sBHs and the stars (Eqs. \ref{eq:nBH} and \ref{eq:nstar}) ensure a global zero-flux solution and thus a smooth transition between the star-dominated region, at $r\gtrsim\RI$, and the sBH-dominated one, at $r\lesssim\RI$ \citep[for further details, see][]{Linial_2022}.

A main outcome of our model is the broken-power-law density distribution of sBHs (Eq. \ref{eq:nBH}), rather than a single power law, where the number density's exponent is $\gamma=7/4$, for a BW profile, or $\gamma\approx2$, for a strongly segregated cusp. 
In both of the latter distributions, the majority of the sBHs are around $r\sim R_h$. Since EMRIs originate from $r\sim R_c\ll R_h$ (see Eq. \ref{eq:Rc}), these models predict that plunges are more common than EMRIs, roughly by a factor of $R_h/R_c\sim30$ \citep{Alexander_2017}. 
However, in our steady-state distribution, the sBHs are concentrated at $r\sim\RI$ (Eq. \ref{eq:RI}), leading to a reduced rate of plunges, compared to the single-power-law models. Furthermore, we predict that EMRIs become more common than plunges if $\RI\leq R_c$, i.e., $\fbh\leq f_\bullet^c$.

Considering the eccentricity distribution, we find that at the mostly bound orbit, the EMRI eccentricity follows a narrow distribution, peaking at small residual eccentricity, $e\lesssim0.1$, which is compatible with previous estimates \citep[e.g.,][]{Babak_17}.

Regarding the eccentricity evolution throughout the merger, we adopt a simplified approach. While other studies \citep[e.g.,][] {Bonetti_2020,Pozzoli_2023} account for the broadband GW emission along eccentric orbits, we calculate the GW strain assuming circular orbits. This is motivated by the typically low eccentricities at the vicinity of the SMBH, as discussed above. Moreover, emission from tightly bound, nearly circular orbits dominates over that from eccentric orbits at the same frequency \citep{Peters_1963}. Thus, our method captures the dominant GW signal, though it neglects the extended, lower-frequency tail associated with the circularization of eccentric orbits during the merger.

Previous estimations of the number of sources detectable by LISA span 3 orders of magnitude, from a few to thousands of EMRIs per year \citep{Barack_2004,Gair_2004,Mapelli_2012,Babak_17,Bonetti_2020,Pozzoli_2023}.
\cite{Babak_17} conducted a comprehensive study on the influence of various parameters on the number of detected EMRIs, which has been adopted in subsequent works \citep[e.g.,][]{Bonetti_2020,Pozzoli_2023}.
Their analysis showed that a positive SMBH mass function slope (see Eq. \ref{eq:rhoSMBH}) results in a low detection rate, below a few tens of EMRIs per year.

When considering the SMBH mass function adopted here, with its negative slope, the key factor leading to the broad range in the predicted number of sources is the EMRI-to-plunge ratio. While \cite{Babak_17} treated it as a free parameter, in our model the EMRI-to-plunge ratio is fully determined by the dynamics and the sBH number fraction, $\fbh$. 
For a typical value of $\fbh\approx10^{-3}$, we find roughly $2.5$ plunges per EMRI (see Table \ref{tab:1}), contrary to the commonly assumed values of $\approx10-100$ \citep[e.g.,][]{Alexander_2017,Babak_17}.
This difference arises from the stronger mass segregation in our model, which leads to a reduced number of plunges.
Even for a mass fraction as large as $\fbh\approx10^{-2}$, our model predicts only about $20$ more plunges than EMRIs.

Additionally, \cite{Babak_17} consider the effects of the SMBH spin, the $\Mbh-\sigma_h$ relation, and cusp erosion. These factors typically introduce order unity changes in the expected number of sources and are not included in our analysis.

We note that the intrinsic EMRI rate per galaxy used by \cite{Babak_17} is based on the Fokker-Planck calculations of \cite{Amaro_Seoane_2011} for a strongly segregated cusp. 
\cite{Amaro_Seoane_2011} predict several hundreds EMRIs per Gyr, for a Milky Way-like galaxy, a value that is comparable to our result (Eq. \ref{eq:rateE}). Furthermore, their scaling relation, $\Gamma_{\rm EMRI}\propto\Mbh^{-0.19}$, is similar to our analytical result: $\Gamma_{\rm EMRI}\propto\Mbh^{-0.25}$.

Overall, when comparing our estimated detection rates with those of \cite{Babak_17}, \cite{Bonetti_2020}, and \cite{Pozzoli_2023}, we observe that our predictions consistently fall above their results for the fiducial model (M1) and below their results for the ``optimistic'' model (M12), despite the differences in model assumptions and calculation methods. 
This trend aligns with the fact that our model analytically finds a lower number of plunges per EMRI than the M1 model and a higher number than M12 model, highlighting the significant influence of this parameter \citep[as previously noted by][]{Babak_17}.
In our analysis, this is not a free parameter but calculated from the dynamics, as discussed in Section \ref{sec:3}.

Last, our GWB characteristic strain scaling, $h_{c,\rm GWB}\propto f^{-1.26}$, is steeper than the $h_{c,\rm GWB}\propto f^{-1}$ obtained by \cite{Bonetti_2020}.
Additionally, We estimate the GWB SNR \citep[following][]{Bonetti_2020,Pozzoli_2023}, yielding $SNR_{\rm GWB}\approx1500$, which is also between the values calculated by \cite{Bonetti_2020} for the M1 and M12 models.
Note that \cite{Pozzoli_2023} found that the detection rates and GWB SNR are lower by a factor of a few, compared to the results presented here and those from \cite{Babak_17} and \cite{Bonetti_2020}. For a detailed discussion, see Section V.a of \cite{Pozzoli_2023}.

It is worth noting that the GWB calculations in \cite{Bonetti_2020} and \cite{Pozzoli_2023} follow a different approach than the one applied here. They employ an iterative estimation of the GWB energy density \citep[following][]{Phinney_2001}, including the summation of higher GW harmonics and the subtraction of resolvable sources.
Additionally, beyond the different model assumptions discussed above, each of the cited works, including our own, adopted slightly different sensitivity curves for LISA and waveform models that also contribute to the deviations in the results, if expressed in terms of SNR.

\section{Summary} \label{sec:sum}
In this work, we study the dynamics around SMBHs, focusing on the formation of EMRIs, driven by two-body scattering, and its implications for LISA observations.
We combine the steady-state solution, due to two-body scattering, as derived by \cite{Linial_2022}, with a GW-induced dissipation. This leads to a schematic division of the phase space into a scattering-dominated region and a GW-dominated one, as depicted in Fig. (\ref{fig:1}). Each region is further characterized by a broken-power-law distribution, as given in Eq. (\ref{eq:nBH}).

Using the above result, we estimate the formation rates of EMRIs and plunges, per SMBH, and the EMRI eccentricity distribution, which are qualitatively comparable to previous results in the literature (see Section \ref{sec:comp}), although derived from a different model for the steady-state distribution in NSCs. 

However, our model differs from previous estimations in its prediction of the EMRI-to-plunge ratio; while it is usually assumed that plunges are more prevalent than EMRIs, our model predicts that if the sBH population is scarce, i.e, its number fraction $\fbh<f_\bullet^c\simeq 4.5\times10^{-4}$, the EMRI rate will exceed the plunge rate. For more massive sBHs, the critical number density, $f_\bullet^c$, is smaller and the EMRI-to-plunge ratio increases more rapidly, as evident in Fig. (\ref{fig:3}).

We further estimate the GWB induced by a cosmological population of EMRIs, which hampers the LISA sensitivity in the range $1-5\ \rm mHz$, by up to a factor of $\approx2$ around $2.5\ \rm mHz$. Accordingly, we estimate that during a $4$ yr mission, LISA will detect approximately $2\times 10^3$ resolvable sources, with SNR greater than $\left.\rm {SNR}\right|_{\rm th}=20$.

\begin{acknowledgments}
This research was partially supported by an ISF grant, an NSF/BSF grant, and an MOS grant. B.R. acknowledges support from the Milner Foundation. I.L. acknowledges support from a Rothschild Fellowship, The Gruber Foundation, and a Simons Investigator grant, 827103. K.K. gratefully acknowledges support from the Israel Science Foundation (Individual Research grant 2565/19).
\end{acknowledgments}

\newpage
\appendix
\section{Formation Rates of EMRIs \& Plunges} \label{app:a}
The distribution of a population of sBHs with a given mass $\mbh$, i.e., the equivalent to Eq. (\ref{eq:nBH}), can be derived following the same steps described in Section (\ref{sec:2}), in a straightforward yet algebraically cumbersome manner. 
Here, we present explicitly the expressions for the formation rates of EMRIs and plunges. 
As in Section (\ref{sec:3}), we assume $\RGW\ll\RI,R_c\ll R_h$. 
To simplify the notation, we denote $\widetilde{m}_{\bullet}=\mbh/\mstar$.

Considering the EMRI rate, from Eq. (\ref{eq:rateE}), we get
\begin{equation}
    \Gamma_{\rm EMRI}\simeq1050\widetilde{m}^{-3/5}_\bullet\left(\frac{\Mbh}{\Mmw}\right)^{-1/4}\left\{\def\arraystretch{1}\begin{tabular}{@{}l@{\quad}l@{}}
        $1$ & $\fbh> f^c_\bullet$ \\
    $75\left(\fbh\widetilde{m}_{\bullet}\right)^{4/5}\left[1+\frac{4}{\widetilde{m}_{\bullet}-5}\left(1-0.013^{(5-\widetilde{m}_{\bullet})/4}\left(\fbh \widetilde{m}_{\bullet}\right)^{\widetilde{m}_{\bullet}/5-1}\right)\right]$ & $\fbh<f^c_\bullet$ \\
    
\end{tabular}\right. .
\end{equation}

The plunge rate, Eq. (\ref{eq:rateP}), is given by
\begin{equation}
    \Gamma_{\rm Plunge}\simeq78216\left(\frac{\Mmw}{\Mbh}\right)^{1/4}\left\{\def\arraystretch{1}\begin{tabular}{@{}l@{\quad}l@{}}   
       $\fbh^{4/5}\widetilde{m}_{\bullet}^{1/5}\left[1+\frac{4}{\widetilde{m}_{\bullet}-5}-0.01\left(\fbh\widetilde{m}_{\bullet}\right)^{-4/5}\right]$ & $\fbh> f^c_\bullet$ \\
     $4\left(\fbh\widetilde{m}_{\bullet}\right)^{(\widetilde{m}_{\bullet}-1)/5}\frac{0.013^{(5-\widetilde{m}_{\bullet})/4}}{\widetilde{m}_{\bullet}^{3/5}\left(\widetilde{m}_{\bullet}-5\right)}$ & $\fbh<f^c_\bullet$ 
\end{tabular}\right. .
\end{equation}

Finally, using the above results, we get the EMRI-to-plunge ratio in the limiting cases:
\begin{equation}\label{eq:E2Ptot}
\frac{\Gamma_{\rm EMRI}}{\Gamma_{\rm Plunge}}\simeq \left\{\def\arraystretch{1}\begin{tabular}{@{}l@{\quad}l@{}}
        $55\left(\widetilde{m}_{\bullet}-1\right)0.013^{\widetilde{m}_{\bullet}/4}\left(\fbh\widetilde{m}_{\bullet}\right)^{1-\widetilde{m}_{\bullet}/5}$ & $\fbh\ll f^c_\bullet$ \\
        $0.01\frac{\widetilde{m}_{\bullet}-5}{\widetilde{m}_{\bullet}-1}\left(\fbh\widetilde{m}_{\bullet}\right)^{-4/5}$ & $\fbh\gg f^c_\bullet$ \\
\end{tabular}\right. .
\end{equation}
\section{GW Emission from Circular Orbits - GR Corrections} \label{app:c}
We estimate the GR strong-field correction to the GW emission under the adiabatic approximation \citep[for further details, see][]{Theocharis_1993,Hughes_2000,Hughes_2005}. 
Namely, we treat the sBH as a test particle that slowly descends from one circular geodesic, with radius $R$, to another. Thus, the radial velocity can be determined by the orbital-averaged power emitted in GWs, which we determine using the semianalytical method introduced by \cite{Rom_2021}.

Under these assumptions, the GR correction introduces two modifications to the weak-field approximation \citep{Peters_1964}: (a) an increased radiated power near the ISCO, by roughly $\sim15\%$; and (b) a shallower effective potential at the ISCO vicinity \citep{Gravitation_1973}, 
which shortens the merger time \citep{Ori_2000,Buonanno_2000}.

Following \cite{Finn_2000}, we define the GR corrections relative to the weak-field approximation.
Therefore, the GR corrections to the GWs' strain, $h$, and characteristic time for changing the orbital frequency, $f/\dot{f}$ (Eq. \ref{eq:st3}), are given by
\begin{equation}\label{eq:GR1}
    \varepsilon_h=\sqrt{\frac{\dot{E}_{\infty,2}}{\left.\dot{E}\right|_{Q}}},
\end{equation}
\begin{equation}\label{eq:GR3}
    \varepsilon_f=\frac{f/\dot{f}}{\left.f/\dot{f}\right|_{Q}}=\frac{R/R_s-3}{2\left(R/R_s-3/2\right)\left(R/R_s-1\right)}\frac{\left.\dot{E}/E\right|_{Q}}{\dot{E}_\infty/E},
\end{equation}
where $\dot{E}_\infty$ is the total power radiated to infinity, $\dot{E}_{\infty,2}$ is the power radiated to infinity in the quadrupole mode, and $\left.\dot{E}\right|_{Q}=\left(\mbh/M\right)^{2}\left(R_s/R\right)^{5}/5$ and $\left.E\right|_{Q}=-G\Mbh\mbh/\left(2R\right)$ are the weak-field estimates. 

Combining the above results, the correction to the characteristic strain for evolving sources is
\begin{equation}\label{eq:GR4}
\begin{aligned}
    \varepsilon_{h_c}&=\varepsilon_h\sqrt{\varepsilon_f}=\left[\frac{R/R_s}{\left(R/R_s-3/2\right)^3}\right]^{1/4}\sqrt{\left(\frac{R}{R_s}-3\right)\frac{\dot{E}_{\infty,2}}{\dot{E}_{\infty}}}\\
    &\simeq\left(1-\frac{\xi_f}{2}\right)^{-3/4}\sqrt{\left(1-\xi_f\right)\left(1-\frac{\alpha}{3}\xi_f\right)},
\end{aligned}
\end{equation}
where $\xi_f=\left(f/f_{\rm ISCO}\right)^{2/3}$ and $f_{\rm ISCO}$ is twice the orbital frequency at the ISCO. Following \cite{Rom_2021}, we approximate $\frac{\dot{E}_{\infty,2}}{\dot{E}_{\infty}}\simeq 1-\alpha \left(R_s/R\right)$. Numerically, we find that for $\alpha=(4/5)^2$ our approximation agrees with the numerical result, with a relative error less than $0.5\%$.

Considering sources at the local Universe (Section \ref{sec:An}), the effect of the GR corrections can be estimated as follows.
For evolving sources, i.e., $f>f_s$, the contribution of the most massive SMBH, $\Mbh=M_{\max}$ is suppressed due to the reduced merger time near the ISCO (Eq. \ref{eq:GR3}). 
This can be taken into account effectively by integrating up to a smaller maximal SMBH mass, $\Upsilon\times M_{\max}$, where 
$\Upsilon\simeq \left\{\def\arraystretch{0.75}\begin{tabular}{@{}l@{\quad}l@{}}
    $0.5$ & $\fbh/f^c_\bullet\lesssim0.6$ \\
       $0.65\times\min\left\{1,\left(\fbh/f_\bullet^c\right)^{3/5}\right\}$ & $\fbh/f^c_\bullet\gtrsim 0.6$ \\
\end{tabular}\right.$.
Thus, the effective maximal SMBH mass corresponds to the SMBH for which the sBH orbits at $R\approx5R_s$ or at $R=\RII$. The former is relevant for $\fbh\lesssim0.6f_\bullet^c$, where $\RII\gtrsim5R_s$, while the latter is valid at larger values of $\fbh$, such that $4R_s\leq\RII\lesssim5R_s$.
This modification translates to an effective parameter $\varepsilon\approx0.82$ in Eq. (\ref{eq:f2}).

On the other hand, for steady sources at low frequencies, $f<f_s$, the reduced merger time is relevant only at the immediate vicinity of the ISCO, as $M_{\rm max}$ is larger at low frequencies and therefore its impact is subdominant. 
In this case, we find that the effective GR correction is mild; the full numerical results are well captured by taking $\varepsilon=0.93$. This can be attributed to the reduced emission power in the $m=2$ mode, resulting in a slightly weaker strain, $\varepsilon\approx\varepsilon_h(R_{\rm ISCO})$.

\section{Cosmological model} \label{app:b}
We assume a standard flat $\Lambda\rm{CDM}$ model, where $\Omega_\Lambda=0.68$, $\Omega_M=0.32$, and $H_0=67\ \rm km\ s^{-1} Mpc^{-1}$ are the dark energy density, matter density, and Hubble constant, respectively \citep{Planck_2020}.

The luminosity distance, $d_L(z)$, and the comoving volume, $\dd V_c/\dd z$, are given by
\begin{equation} \label{eq:dl}
    d_L(z)=\left(1+z\right)\frac{c}{H_0}\int_0^z \frac{\dd \tilde{z}}{\varrho(1+\tilde{z})}=\frac{c}{H_0}\eta(1+z),
\end{equation}
\begin{equation}\label{eq:dV}
    \frac{\dd V_c}{\dd z}=4\pi \left(\frac{c}{H_0}\right)^3\frac{\left(\int_0^z \frac{
    \dd\tilde{z}}{\varrho(1+\tilde{z})}\right)^2}{\varrho(1+z)}=4\pi\left(\frac{c}{H_0}\right)^3\frac{\left[\eta(1+z)/\left(1+z\right)\right]^2}{\varrho(1+z)},
\end{equation}
where $\varrho(x)=\sqrt{\Omega_\Lambda+\Omega_M x^3}$, and
\begin{equation}
    \eta(x)=\frac{x}{\sqrt{\Omega_\Lambda}}\left[x\widetilde{F}\left(-\frac{\Omega_M}{\Omega_\Lambda}x^3\right)-\widetilde{F}\left(-\frac{\Omega_M}{\Omega_\Lambda}\right)\right],
\end{equation}
with $\widetilde{F}\left(x\right)={}_2{\rm{F}}_1\left(1/3,1/2;4/3;x\right)$, and ${}_2{\rm{F}}_1\left(a,b;c;x\right)$ is the hypergeometric function.

\bibliography{main}{}
\bibliographystyle{aasjournal}

\end{document}